# Community Detection Using Revised Medoid-Shift Based on KNN

Jiakang Li[1], Xiaokang Peng[1], Jie Hou[1], Wei Ke[2], Yonggang Lu[1](✉)

[2]Gansu New Vispower Technology Co. Ltd, No. 1689 Yanbei Road, Lanzhou, Gansu 730000, CHINA

✉Corresponding author: ylu@lzu.edu.cn

**Abstract.** Community detection becomes an important problem with the booming of social networks. The Medoid-Shift algorithm preserves the benefits of Mean-Shift and can be applied to problems based on distance matrix, such as community detection. One drawback of the Medoid-Shift algorithm is that there may be no data points within the neighborhood region defined by a distance parameter. To deal with the community detection problem better, a new algorithm called Revised Medoid-Shift (RMS) in this work is thus proposed. During the process of finding the next medoid, the RMS algorithm is based on a neighborhood defined by KNN, while the original Medoid-Shift is based on a neighborhood defined by a distance parameter. Since the neighborhood defined by KNN is more stable than the one defined by the distance parameter in terms of the number of data points within the neighborhood, the RMS algorithm may converge more smoothly. In the RMS method, each of the data points is shifted towards a medoid within the neighborhood defined by KNN. After the iterative process of shifting, each of the data point converges into a cluster center, and the data points converging into the same center are grouped into the same cluster. The RMS algorithm is tested on two kinds of datasets including community datasets with known ground truth partition and community datasets without ground truth partition respectively. The experiment results show sthat the proposed RMS algorithm generally produces betster results than Medoid-Shift and some state-of-the-art together with most classic community detection algorithms on different kinds of community detection datasets.

## 1 Introduction

Social networks have become ubiquitous in our day-to-day lives through such platfms as Facebook, Twitter, and Instagram. These networks can be modeled as graphs, with nodes representing individuals and edges representing their interconnections. Within these complex graphs, certain subgraphs exhibit particularly high density, where individuals are more closely interconnected than elsewhere. These subgraphs are commonly referred to as communities [1].

In recent years, a plethora of community detection algorithms have been proposed to mine hidden information within networks [2]. These algorithms are generally categorized into two types: overlapping and non-overlapping methods. To address the problem of community detection in network analysis, researchers have employed a variety of approaches, including hierarchical divisive, hierarchical agglomerative, and random walk-based methods [3], among others. To evaluate the performance of these algorithms, researchers have proposed various detection metrics. Among these, modularity [4] is a critical metric used to assess the quality of the community generated by different methods. A higher modularity value indicates better community creation [5]. Modularity measures the degree to which nodes within a community are more densely connected than nodes outside that community. Additionally, Normalized Mutual Information (NMI) [6] is a crucial evaluation metric when ground truth partitions exist for the dataset. The higher the NMI, the better the match with the ground truth partition.

Community detection poses a formidable challenge due to the complexity and scale of network structures. Notably, contemporary complex networks primarily comprise graphs, a form of structured data lacking coordinates that precludes the direct utilization of coordinate-based algorithms, such as the Mean-Shift algorithm, for community detection [7]. While the Medoid-Shift algorithm proposed subsequently can address distance matrix-based issues, its application to community detection problems remains largely unexplored. Additionally, the Medoid-Shift algorithm may encounter a critical challenge of no data points existing within the neighborhood region defined by its distance parameter, leading to suboptimal performance on community detection problems.

Therefore, to address the challenges above, this paper has proposed a new community detection algorithm named RMS, which extracts the characteristics from both $k$ nearest neighbor (KNN) [9] and Medoid-Shift while focusing on detecting the non-overlapping community. In contrast to the traditional Medoid-Shift algorithm, our proposed method employs a modified approach for determining the neighborhood of a given node. Specifically, we have defined a parameter $k$ for RMS borrowing the idea of KNN. The parameter $k$ is used to define the medoid's neighborhood, rather than using a distance parameter as in the conventional Medoid-Shift algorithm. This modification effectively mitigates the issue of unstable number of data points within the defined neighborhood regoin, which is a known limitation of the original Medoid-Shift method. Moreover, during the shifting of the medoid, the RMS algorithm caculates similarities between each point $p$ in the neighborhood of the current medoid and the KNN of $p$. Because the KNN of $p$ is not related with the current medoid, the stableness of the shifting process is enhanced compared to the original Medoid-Shift method.

The following are the main contributions of this paper:

• Introducing the concept of Medoid-Shift to community detection for the first time.

• By introducing KNN into the Medoid-Shift method, a novel RMS method is proposed for community detection.

• Compared to the original Medoid-Shift method, the proposed method can perform community detection more effectively.

The remaining portions of this paper are organized as follows: Section 2 discusses related works in community detection. Section 3 discusses the process of the RMS algorithm and its data pre-processing in detail. In section 4, we present the experimental results and analyze them in detail. Section 5 discusses the conclusion and future work.

## 2 Related Work

The research on uncovering the community structure of the real social network has been a hot research topic, which has also spawned many community detection algorithms. This section first introduces classical algorithms, KNN-based algorithms, and distance matrix-based algorithms. Then it introduces Medoid-Shift.

### 2.1 Classical Algorithms in Community Detection

In 2004, Newman and Girvan developed the GN [10] algorithm which is a well-known method for discovering communities. The algorithm employs divisive hierarchical clustering and iteratively removes edges with the highest betweenness score to partition the network into subgroups. In 2010, Louvain [11] introduced a community detection algorithm that focuses on optimizing modularity. The algorithm aims to maximize the modularity of the entire network through an iterative process of optimizing the partition of the network into communities. Besides these two, there have been numerous community detection algorithms proposed during the past two decades, each with its characteristics and advantages. For instance, module-based optimization [12] algorithm, spectral clustering [13] algorithm, hierarchical clustering [11] algorithm, label propagation [14] algorithm, and information theory-based algorithm [15], which have become classical algorithms in community detection.

### 2.2 KNN-Based Algorithms

Based on the idea of the KNN method, there are numerous community detection algorithms. Dong and Sarem have proposed an overlapping algorithm called NOCD [16] based on KNN. The improved KNN algorithm brings a new approach to the table by replacing network distance with similarity.Jia and Li [17] introduced KNN-enhance, an uncomplicated and adaptable community detection method incorporating node attribute enhancement. The approach utilizes the KNN graph of node attributes to enhance the community structures of network by reducing the effects of sparsity and noise in the original network.

### 2.3 Distance/Similarity Matrix-Based Algorithms

The distance and similarity are opposite in denoting the closeness between nodes. But they can transform into each other to some extent. The following content introduces both distance matrix and similarity matrix-based algorithms. The K-medoids algorithm [18], an approach similar to the K-means, could also be based on a distance matrix, and can effectively identify non-spherical clusters and determine the optimal amount of clusters automatically. Zhang and Jin have proposed an algorithm called CFSFDP [19]. It defines a trust-based distance measure and quantifies user relationships in social networks as distance matrices using kernel density estimation. Sona and Asgarali [20] proposed an affinity propagation-based method with adaptive similarity to community detection called APAS. It constructs an initial similarity matrix based on some similarity metric between nodes. Besides that, it also applies the affinity propagation algorithm to the similarity matrix to find the exemplars, or representatives, of the network.

### 2.4 Medoid-Shift Algorithm

Derived from the idea of Mean-Shift, Medoid-Shift is very similar with Mean-Shift in the theory and process. They both calculate the shift toward regions of greater data density, find the cluster centers by iteration, and calculate the number of clusters automatically. The biggest difference, as shown in the Fig. 1, is that Mean-Shift shifts to a location according to the Mean-Shift vector, while this algorithm shifts to a certain point in the neighborhood. Furthermore, Medoid-Shift can be directly applied to distance-based community detection problems, but Mean-Shift can not.

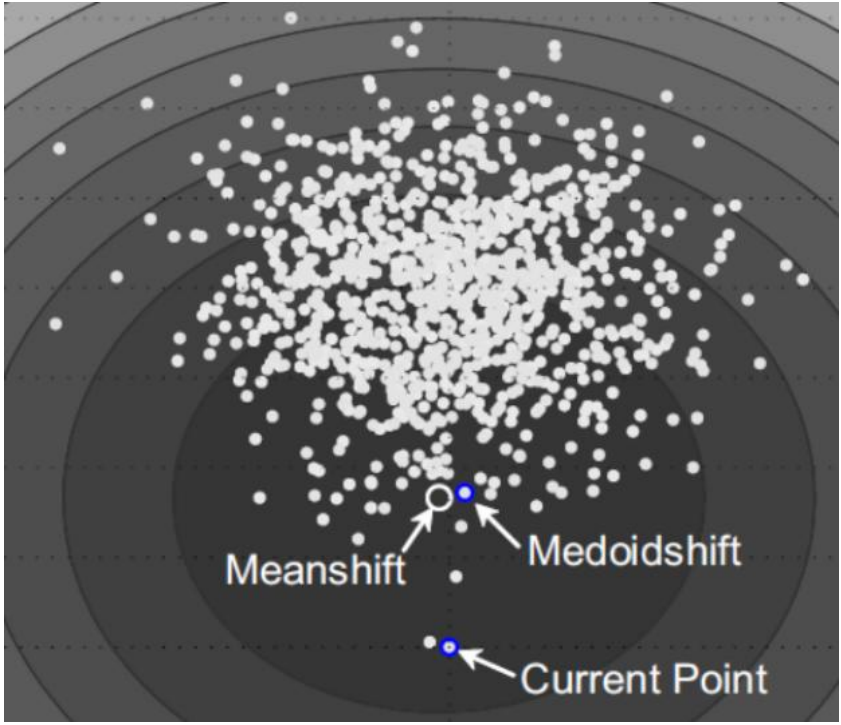

**Fig. 1.** For the current point in the shifting, Medoid-Shift chooses the data point to shift, while Mean-Shift chooses the location to shift [8].

**Brief Introduction of Medoid-Shift Core Algorithm.** Given an $N \times N$ symmetric matrix $\mathbf{D}(i, j)$ which is the distance between $i$ and $j$ starting from the point $i$, an index of point $j$ is calculated as follows:

$$S(i,j) = \sum_{k=1}^{N} D(j,k)\phi(D(i,k))  \quad (1)$$

The next point to shift for $i$ in the shifting is point $j$ with the minimum value in S($i$, $j$), 1<=$j$<=$n$. By iteratively computing the next point from all the current points, tree traversal can be used to find the unique roots for all the points [8]. The number of clusters is the number of unique roots, and the label of each point can be derived directly from its corresponding unique root.

## 3    The Proposed RMS Algorithm

The RMS algorithm proposed in this study differs from the Medoid-Shift algorithm in several aspects. Firstly, the neighborhood in Medoid-Shift is defined by a distance parameter, whereas in RMS, it is defined by KNN. Secondly, in Medoid-Shift, distances between all the points in the current medoid's neighborhood are calculated when selecting the next point, while in RMS, similarities between each point *p* in the neighborhood of the current medoid and the KNN of p are computed for this purpose.

In the remainder of section 3, the proposed RMS data preprocessing method is firstly introduced. Then we introduce the core algorithm of RMS and analyze the advantages of RMS compared to the Medoid-Shift algorithm.

### 3.1 The Definition of Graph in Social Network.

Given a social network *N*, it can be denoted as a graph G(*V*, *E*), where $u \in V$ is a node, $e \in V$ is another node, $(u, e) \in E$ is an edge that measures some form of relationship (intimacy) between *u* and *e*. The number of edges and nodes are denoted by |*E*| and |*V*| respectively. It's worth mentioning that, since RMS only considers whether there is a connection between the two endpoints rather than the direction, directed graph is thus coverted to an undirected graphs by adding up the weights of the directed edges between every two endpoints.

### 3.2 Data Preprocessing of the RMS Algorithm

In this study, a similarity matrix is utilized instead of a distance matrix to represent the weights of the graph. Specifically, the similarity matrix, denoted as SimM(*i*, *j*), captures the similarity between nodes *i* and *j*, where NN(*i*) represents the K-nearest neighbors of node *i*, and N(*i*) represents the list of points connected to node *i*. The entire community is modeled as an undirected graph with a similarity matrix, where the relevance of nodes *i* and *j* is denoted by the value of coordinate(*i*, *j*) in the matrix. The higher the value of the relevance, the more likely it is that nodes *i* and *j* belong to the same community. For an undirected graph, the distance matrix is typically an upper or lower triangular matrix.

In particular, when dealing with weighted graphs, the weight between two points is typically represented by their similarity value. However, for unweighted graphs, it is not appropriate to use the same weight scheme as in the case of weighted graphs. To address this issue, a novel method is proposed for computing the weights. The similarity between two points *i* and *j* in the unweighted graph is defined as:

$$SimM(i,j) = len(N(i) \cap N(j)) \tag{2}$$

### 3.3 Core Process of the RMS Algorithm

The basic process of the RMS algorithm involves the use of three tools: Finding the KNN and similarity sum for each point, medoid clustering, and label assignment.

**Finding the KNN and Similarity Sum for Each Point.** We define a property "Similarity Sum" represented as DL(*i*) for point *i*, which is the sum of the similarity between the point *i* and its *k* nearest neighbors. The larger the "Similarity Sum" is, the more likely the point will be the medoid.

$$DL(i) = \sum_{p \in NN(i)} SimM(i,p) \tag{8}$$

Before iterating, the "Similarity Sum" is calculated for each data point. Function 1 describes the process for calculating the "Similarity Sum" and KNN for each data point.

**Medoid Clustering.** After the computation of the "Similarity Sum" and KNN, the proposed RMS algorithm initializes all points as the initial medoids. As shown in Algorithm 1, for each medoid $i$ in SetA, the algorithm selects the next medoid from the $k+1$ points which include the point i and its KNN. Specifically, the point with the largest "Similarity Sum" is chosen as the next medoid after point $i$, and is added to the new medoid set called SetB. This process continues for all the points in SetA. After each iteration, the algorithm compares the previous medoid set, SetA, with the newly obtained medoid set, SetB. If the two sets are identical, the iteration stops; otherwise, SetB is assigned to SetA, SetB is cleared up, and a new iteration starts. Once the iteration stops, the points in SetA are returned as the cluster centers, and the next medoid of each point is stored in a list called nextMedoid.

**Label Assignment.** After getting the cluster centers and the next medoid of each point, the data points converging into the same center are grouped into the same cluster.

---

**Function 1** Finding the KNN and Similarity Sum for Each Point

**INPUT**: SimM(1..$|V|$, 1..$|V|$), $k$
**OUTPUT**: DL, NN

**for** $i \leftarrow 1$ to $|V|$ **do**
    NN[$i$] $\leftarrow$ The indices of the $k$ nearest neighbors of point $i$
    DK $\leftarrow$ The $k$ largest values in SimM[$i$,:]
    DL[$i$] $\leftarrow$ sum(DK)
**end for**
**return** DL, NN

---

**Algorithm 1** Medoid Clustering

**INPUT**: SimM(1..$|V|$, 1..$|V|$), $k$
**OUTPUT**: cluster centers, the center points list of final clustering result

DL, NN $\leftarrow$ **Function 1**(SimM, $k$)
SetA $\leftarrow \{1,2, ..., |V|\}$
SetB $\leftarrow \emptyset$
nextMedoid $\leftarrow$ []
**while** true **do**
    **for** each $i$ in SetA **do**
        SetP $\leftarrow \{i\} \cup$ NN($i$)
        temp $\leftarrow$ The point with the largest DL value in the SetP
        nextMedoid[$i$] $\leftarrow$ temp
        SetB.add(temp)
    **end for**
    **if** SetA == SetB **then**
        **for** each $i$ in SetA **do**
            nextMedoid[$i$] $\leftarrow i$
        **end for**
        **break**

        **end if**
        SetA, SetB ← SetB, Ø
**end while**
centers ← SetA
**return** centers, nextMedoid

---

**Algorithm 2** Label Assignment

**INPUT**: nextMedoid
**OUTPUT**: labels, a list contains labels of each point
labels ← [ ]
**for** $i \leftarrow 1$ to $|V|$ **do**
    $m \leftarrow i$
    $k \leftarrow$ nextMedoid[$m$]
    **while** $m \neq k$ **do**
        $m \leftarrow k$
        $k \leftarrow$ nextMedoid[$m$]
    **end while**
    labels[$i$] ← $m$
**end for**
**return** labels

**Complexity analysis.** In the worst case, the complexity can be $|V|^2$. But in most cases, it will be better than $|V|^2$. For the function 1, it is

## 4 Experimental results

Section 4 begins by describing the experimental setup, followed by an introduction of evaluation metrics and an overview of the comparative methods used in the study. The discussion is then divided into two parts: weighted graphs without ground truth, and unweighted graphs with ground truth. For each part, the section provides an introduction to the datasets used, the process of parameter tuning, the results of the comparative experiments, and a detailed discussion.

### 4.1 Overall Experiment Setup

Instead of utilizing the distance matrix, our approach employs the similarity matrix as the input for all algorithms. Notably, the datasets with known ground truth partitions consist solely of unweighted graphs, while the datasets without ground truth partitions consist solely of weighted graphs. This is because ground truth partitions can only be collected for the unweighted graphs. To evaluate the results of the experiments on the datasets without ground truth partitions, we use modularity, while normalized mutual information (NMI) is used for the datasets with ground truth partitions. To optimize the modularity and NMI values in the proposed RMS approach, a parameter, k, is defined. We set a range of k values within the datasets to tune and obtain the best results.

    Through experimentation, we find that our proposed approach achieves a better overall modularity value for data without ground truth partitions, and a better NMI value for data with ground truth

partitions, than other algorithms, including Medoid-Shift, Label Propagation, Infomap algorithm, NNSED, SCD, and Girvan Newman algorithm. To implement Medoid-Shift, we use Φ(D(i, j))= exp(-D(i, j)/2).

### 4.2 Evaluation Metrics.

Similar to other community detection researches, we utilize the following metrics to assess the effectiveness of algorithms.

**Normalized mutual information.** The NMI value serves as an additional assessment parameter for community detection with ground truth partition. Generally, a higher NMI value is indicative of a closer partition to a real partition and is widely accepted.

$$NMI(I,C) = \frac{2 \times I(Y;C)}{H(Y) + H(C)} \tag{3}$$

where $C$ is the result of clustering, and Y is the actual category of data. H(.) is cross entropy and is denoted as:

$$H(X) = -\sum_{i=1}^{|X|} P(i) \log_2 P(i) \tag{4}$$

where $I(Y; C)$ is mutual information and is denoted as:

$$I(Y;C) = H(Y) - H(Y|C) \tag{5}$$

**Modularity $Q$.** The accuracy of community detection in a complex network is conventionally evaluated by the modularity function $Q$, which is widely recognized as a standard. The modularity is usually denoted as follows:

$$Q = \frac{1}{2m} \sum_{ij} (A_{ij} - \frac{k_i k_j}{2m}) \sigma(i,j) \tag{6}$$

where $A$ is the adjacency matrix, $m$ is number of edges in the graph, $\delta(i, j) = 1$ if both point $i$ and point $j$ are in the same community and 0 otherwise. However, this research relies heavily on weight calculation, while an unweighted graph has no weight. To solve this problem, we count the weight of every edge as 1 in an unweighted graph according toTo address this issue. The modularity of weighted graphs is calculated using the following formula:

$$Q = \sum_{1}^{k} (\frac{e_{kk}}{m} - (\frac{d_k}{2m})^2) \tag{7}$$

where $d_k$ is sum of the degrees of all nodes that are in the community $k$, $e_{kk}$ is the sum of the weights of all edges in community $k$.

### 4.3 Comparative Methods

This study employs both classical and state-of-the-art algorithms that focus on non-overlapping communities as comparative experiments on both weighted and unweighted graph data. Compared to other methods, the RMS approach demonstrats its advantages by achieving higher modularity scores in weighted graphs and higher normalized mutual information values than most classical and some state-of-the-art methods. The algorithms used for comparative experimentation in this article are listed below:

**Label Propagation.** The underlying assumption of label propagation is that a node's label can be reconstructed in a linear fashion using the labels of its neighbors, allowing for smooth propagation of the node's label to its neighboring nodes. This process is repeated until all nodes attain stable labels.

**SCD [21].** Based on network node embedding, this algorithm for community detection evaluates the quality of community detection by measuring the Silhouette contour coefficient and optimizing it.

**GEMSEC [22].** By utilizing a random walk strategy, it represents nodes in an abstract vector space and simultaneously conducts node clustering and embedding. As a result, it exhibits notable robustness and community partitioning capabilities.

**EdMot [23] .** It captures the high-order graph's properties at the level of individual nodes and edges. Furthermore, based on current shortcuts for dealing with a high-order network, it presents an Edge improvement strategy for the Motif-aware community discovery method (known as EdMot).

**Infomap.** Infomap identifies communities by minimizing the map equation, which seeks to locate communities of nodes with the shortest path connections. To achieve this goal, Infomap uses a random walker to detect paths between nodes, and each step taken by the walker is marked with a sequence of codes. The random walker halts its exploration when it can no longer reduce the length of the coded paths.

**Girvan Newman.** The method of identifying edges with the highest betweenness was employed by the authors to propose a divisive algorithm. Betweenness centrality, which was introduced by Freeman, estimates the shortest path and assigns a much higher score to edges between communities compared to those within communities. The algorithm utilizes this concept to iteratively remove the edge with the highest betweenness score until the graph is completely divided into isolated nodes.

### 4.4 Experimental Results for the Datasets without Ground Truth Partition

Given the datasets have no ground truth partition, NMI is not suitable for evaluation. Thus modularity is adopted to assess algorithms in this section.

**Dataset Description.** The performance of the RMS algorithm are assessed on five real-world datasets, including Cell Phone Calls [24], Enron Email [25], Les Miserable network [26], and US airports [27]. We have not used datasets that are larger than these, because they are computationally hard for RMS to solve.

*Enron Email Dataset.* Consisting of 184 nodes and 125,409 edges, it is an email dataset publicly released by the U.S. Department of Justice. As needs mentioning, enron email social network is such a high-order network that we can't directly use the original dataset. Thus we have used the dataset created by Austin R. Benson [25] instead, which is a high-order social network consisting of many simplices. Simplices means a set of nodes during a period of time. When projected to graph form, the dataset has 143 nodes and 1800 edges in total.

*Cell Phones Records Dataset.* The dataset consists of cell phone call records made by Paraiso movement members over a ten-day period in June 2006. This data is used to build a network in which each node represents a distinct cell phone and edges are formed whenever a phone call between two cell phones occurs. Details of the date and time of the phone calls are logged for each edge, with a total of 400 nodes in the network. Similar to the Enron Email Dataset, weights of edges between two endpoints from different periods are added to one value in order to mitigate the impacts of the sequential substance of the network.

*Les Miserable Dataset.* Knuth created the Les Miserable network by analyzing the interconnections of the primary characters in Les Miserables. With 77 nodes and 508 edges, each node signifies a character, and every edge implies the co-occurrence of the associated characters in one or more scenes.

*US Airports Dataset.* Consisting of 1574 vertices and 28,236 edges, the network is a representation of flights that occurred within the United States during 2010. A graph with numerous edges highlights the interconnections between various airports across the country.

**Implementation.** We have implemented the proposed RMS algorithm and other comparative methods using Python. The comparative methods include Infomap, GEMSEC, Girvan Newman, SCD, and Edmot. Some of the algorithms' source codes are publicly available in Python packages such as karateclub [28] and networkx. All experiments are conducted on a Windows machine with 200GB of memory.

**Parameter Tuning.** To achieve the best scores for modularity, algorithms including RMS, Girvan Newman, GEMSEC, and Medoid-Shift require parameter tuning. In Girvan Newman, edge having the highest edge betweenness centrality value than other edges is deleted in every iteration, with each deletion of the edge denoted as m in Figure 2. For the GEMSEC algorithm, combinations of community partitions are calculated, and the n value in Figure 2 represent the number of communities in one partition by the GEMSEC algorithm. For the RMS algorithm, the parameter *k* represented *k* nearest nodes to a medoid. In order to optimize the RMS's results on datasets, the *k* was gradually increasing until the experiment could no longer be run. For the Medoid-Shift algorithm, the parameter *t* denotes the number of iteration time, the actual radius selected for optimizing Medoid-Shift is shown in Equation 8.

$$radius = (t - 1) * (ma + 0.1) / 30 \qquad (8)$$

where ma denotes the max similarity value in the similarity matrix, and the value of t ranges from 1 to 30.

For parameter tuning process in Fig 2, it can be observed that the proposed RMS algorithm and Medoid-Shift require relatively fewer iterative times to achieve the best results compared to other methods, as shown in the figures. Notably, all the modularity scores first ascend until they reach their global optimal values and then descend. The RMS algorithm outperforms Medoid-Shift in terms of modularity scores on all of the datasets as the curve for RMS is generally higher than Medoid-Shift. The settings of parameter *k, t, m* and *n* for achieving the optiaml values are presented in Table 1.

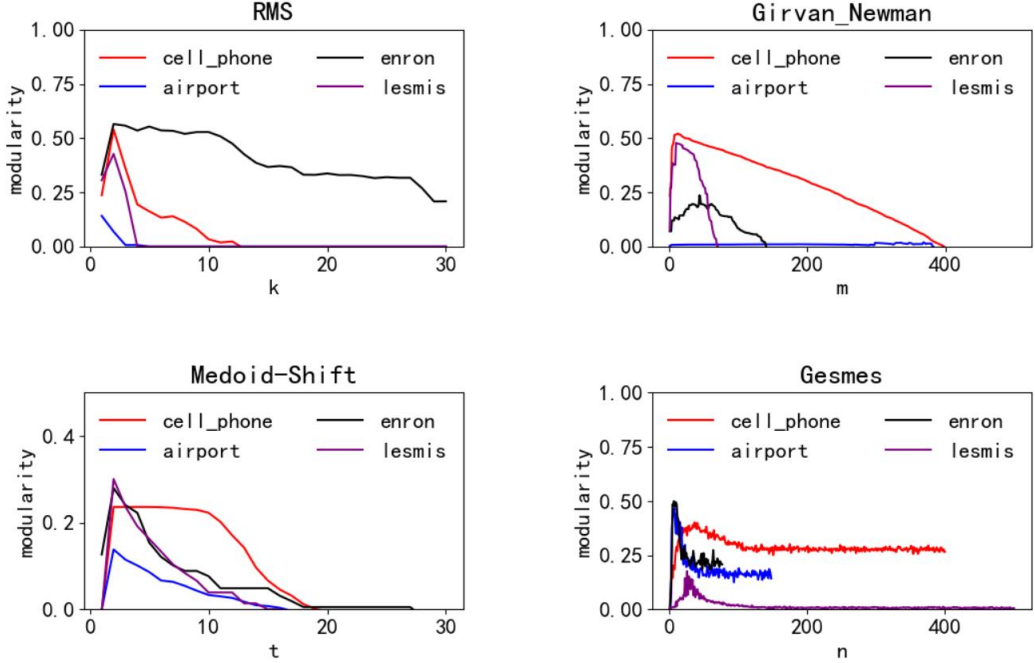

**Fig. 2.** The different modularity scores under different parameter settings for RMS, Girvan Newman, Medoid-Shift, GESMES on datasets without ground truth partition.

Table 1. The settings of parameters *k, t, m* and *n* for optiaml values.

| Parameter | Cell Phones | Enron Email | Lesmis | USA Airport |
|---|---|---|---|---|
| *k* | 2 | 2 | 2 | 1 |
| *t* (radius) | 2 (13.8665) | 2 (2.8275) | 2 (1.0724) | 2 (7.7758) |
| *m* | 15 | 46 | 12 | 300 |
| *n* | 36 | 7 | 6 | 25 |

**Experiment Results and Discussion.** The results of modularity for the Medoid-Shift, RMS, Label Propagation algorithm, NNSED, SCD, Infomap, and Girvan Newman algorithm are shown in the Table 2 below. Also, the result value is evaluated using modularity and the numerical value in the parenthesis denotes the number of clusters.

Table 2. The modularity and the number of clusters (in parenthesis) for datasets without ground truth parition.

| Algorithm | Cell Phones | Enron Email | Lesmis | USA Airport |
|---|---|---|---|---|
| GEMSEC | 0.4021 (36) | 0.4682 (7) | 0.5001 (6) | 0.1774 (25) |
| SCD | 0.3010 (170) | 0.5339 (17) | 0.4499 (33) | 0.0681 (259) |
| EdMot | 0.6388 (20) | 0.6356 (7) | 0.5648 (8) | 0.2847 (11) |
| Infomap | 0.5869 (68) | 0.6191 (12) | 0.5580 (15) | 0.0434 (36) |
| Girvan Newman | 0.5208 (15) | 0.2365 (46) | 0.4776 (12) | 0.0169 (300) |
| Medoid-Shift | 0.2365 (248) | 0.2802 (90) | 0.2973 (29) | 0.1385 (281) |
| RMS | 0.5379 (78) | 0.5650 (14) | 0.4271 (7) | 0.1413 (40) |

Based on Table 2, it is evident that the proposed RMS algorithm consistently outperforms most other methods, except for Edmot. For instance, in the Enron Email Dataset, the RMS algorithm improves modularity by 0.33, 0.03, and 0.1 compared to Girvan Newman, SCD, and GEMSEC, respectively. This indicates that the RMS strategy is effective in datasets without ground truth partition. Furthermore, in comparison to the Medoid-Shift algorithm, the RMS algorithm has improved modularity by 0.3, 0.28, 0.13, and 0.01 on four different datasets. It indicates that the RMS algorithm improves modularity based on KNN instead of distance parameter, and in general, it is a successful modification from Medoid-Shift that has adapted to community detection effectively.

It is worth noting that the modularity scores for all methods are relatively low in the USA airports dataset. This can be attributed to the high sparsity of the dataset, which makes it difficult for graph-based methods to effectively explore the underlying graph structure.

### 4.5    Datasets with Ground Truth Partition

Given that the datasets have ground truth partition, NMI is the perfect evaluation metric under such circumstance compared to using modularity. Thus we adopt the NMI to assess all of the algorithm in this section.

**Dataset Description.** We have collected the unweighted datasets with ground truth partition, which are American Football Network, Dolphins Social Network and American Kreb's Book.

*Dolphins Social Network.* The Dolphins Network [26] depicts the social connections between 62 dolphins residing in a community off Doubtful Sound in New Zealand. It is an undirected network that captures frequent associations between the dolphins. The network comprises 159 edges and 62 nodes and is divided into two separate communities.

*American Football Network.* Characterizing the Division I Games during the 2000 season [26] showcases the schedule of the games. This network comprises 115 nodes representing teams, and 613 edges depicting their matchups. The network is subdivided into 12 groups for ground truth partition.

*American Kreb's Book.* Krebs created the American political book [29] network which consists of 105 nodes and 441 edges. The nodes represent books about American politics that can be purchased from Amazon's online bookshop. The edges connecting any two nodes indicate that these books have been bought by the same person. The network has been categorized into three distinct communities - "Liberals", "Neutrals", and "Conservatives" - based on the classification of the political books.

**Parameter Tuning.** The overall parameter tunning setting is similar to the experiment on datasets without ground truth partition, except that the NMI is used instead of the modularity. As is shown in Fig 3, the RMS algorithm still stably outperforms Medoid-Shift. However, what varies from the previous experiments is that there are some anomaly trends as the values ascend with iteration times growing but does not descend. And we later find out that it gradually converges to a value. The setting of parameter k and t for achieving the optiaml values are presented in Table 3:

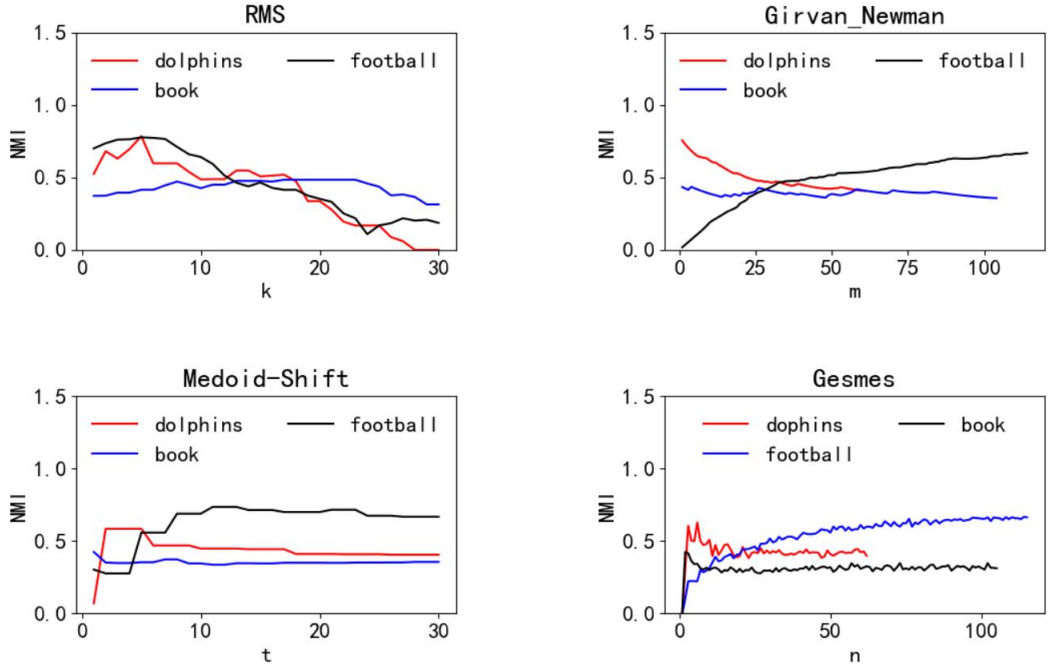

**Fig. 3.** The different NMI values under different parameter settings for RMS, Girvan Newman, Medoid-Shift, Gesmes on datasets with ground truth partition.

Table 3. The setting of parameters k and t for optiaml values.

| Parameter | Dolphins Social Network | American Football Network | American Kreb's book |
|---|---|---|---|
| $k$ | 5 | 5 | 17 |
| $t$ (radius) | 2 (0.2448) | 11 (3.1379) | 1 (0) |
| $m$ | 1 | 113 | 4 |
| $n$ | 6 | 115 | 2 |

**Experiment Result and Discussion.** The results are shown below in the Table 4. The result value is evaluated using NMI and the numerical value in the parenthesis denotes the number of clusters.

Table 4. The NMI and the number of clusters (in parenthesis) for datasets with ground truth.

| Algorithm | Dolphins Social Network | American Football Network | American Kreb's book |
|---|---|---|---|
| GEMSEC | 0.6288 (6) | 0.6690 (115) | 0.4260 (2) |
| SCD | 0.4817 (25) | 0.8664 (14) | 0.3423 (28) |
| EdMot | 0.8353 (4) | 0.7553 (7) | 0.3939 (3) |
| Infomap | 0.7749 (4) | 0.8856 (9) | 0.4378 (6) |
| Girvan Newman | 0.7560 (2) | 0.6684 (114) | 0.4353 (5) |
| Medoid-Shift | 0.5857 (6) | 0.7367 (57) | 0.4258 (2) |
| RMS | 0.7846 (3) | 0.7768 (19) | 0.4840 (2) |

Table 4 shows that the proposed RMS method consistently outperforms most classical and state-of-the-art methods in the Dolphins Social Network and American Kreb's book datasets. In comparison to GEMSEC, SCD, Infomap, and Girvan Newman, the RMS method achieves improved NMI scores of 0.16, 0.3, 0.01, and 0.03, respectively, on the Dolphins Social Network. This suggests that the RMS algorithm is effective even in datasets with a ground truth partition. Additionally, it is important to note that the RMS algorithm performs better than the Medoid-Shift algorithm in the Dolphins Social Network, American Football Network, and American Kreb's book datasets, with improved NMI scores of 0.2, 0.04, and 0.06, respectively. These results demonstrate that the RMS algorithm is a significant improvement over the original Medoid-Shift algorithm.

## 5   Conclusion and Future Work

This paper introduces a novel community detection algorithm called RMS, which builds upon the Medoid-Shift algorithm and incorporates the concept of KNN to map the social network into a distance matrix. This approach addresses the limitations of using the Mean-Shift algorithm directly for community detection and achieves superior performance compared to other traditional algorithms. The study highlights several key insights: (1) The Medoid-Shift algorithm can be extended beyond mode-seeking problems to tackle community detection challenges; (2) The proposed RMS algorithm has the ability to automatically determine the optimal number of communities/clusters; (3) KNN provides a more effective way of defining neighborhood regions compared to using a radius parameter. As a part of our future work, we plan to:

• Incorporate kernel density estimation as part of distance matrix calculation.
• Try an adversarial attack on our algorithm, and see how it performs.
• We will also explore more applications of the RMS clustering algorithm besides community detection.

## Acknowledgments

This work is supported by Gansu Haizhi Characteristic Demonstration Project (No. GSHZTS 2022-2), and the Gansu Provincial Science and Technology Major Special Innovation Consortium Project (Project No. 21ZD3GA002), the name of the innovation consortium is Gansu Province Green and Smart Highway Transportation Innovation Consortium, and the project name is Gansu Province Green and Smart Highway Key Technology Research and Demonstration.